\documentclass[%
 reprint,
 superscriptaddress,
nofootinbib,
 amsmath,amssymb,
 aps,
 prl
]{revtex4-2}

\usepackage{natbib}           
\usepackage{graphicx}
\usepackage{dcolumn}
\usepackage{bm}
\usepackage{hyperref}

\usepackage{soul}
\usepackage{amssymb}
\usepackage{amsmath,bm}
\usepackage{float}
\usepackage[usenames,dvipsnames]{color}
\usepackage[export]{adjustbox}
\usepackage{gensymb}
\usepackage[normalem]{ulem}
\usepackage{latexsym}

\usepackage{soul,xcolor}

\usepackage{multirow}
\graphicspath{{figures/}} 
\usepackage{tikz}
\usepackage[normalem]{ulem}

\newcommand{\cmmnt}[1]{\ignorespaces}

\newcommand{\ket}[1]{\left|  #1 \right\rangle}

\newcommand{\lt}{\ensuremath <}
\newcommand{\gt}{\ensuremath >}

\usepackage{graphicx}
\usepackage{dcolumn}
\usepackage{bm}
\usepackage{hyperref}

\usepackage{amssymb}
\usepackage{amsmath,bm}
\usepackage[percent]{overpic}
\usepackage{mathrsfs}

 \newcommand{\beginsupplement}{%
        \setcounter{table}{0}
        \renewcommand{\thetable}{S\arabic{table}}%
        \setcounter{figure}{0}
        \renewcommand{\thefigure}{S\arabic{figure}}%
     }

\usepackage[normalem]{ulem}
\usepackage{xfrac}
\usepackage{soul}

\date{\today}

\begin{document}

\title{Controlling Rydberg atom-polariton interactions: from exceptional points to fast readout}

\author{Tamara \v{S}umarac}
\thanks{These authors contributed equally}
\affiliation{MIT-Harvard Center for Ultracold Atoms, Massachusetts Institute of Technology, Cambridge, Massachusetts 02139, USA}
\affiliation{Research Laboratory of Electronics, Massachusetts Institute of Technology, Cambridge, Massachusetts 02139, USA}
\affiliation{Department of Physics, Harvard University, Cambridge, Massachusetts 02138, USA}

\author{Emily H. Qiu}
\thanks{These authors contributed equally}
\affiliation{MIT-Harvard Center for Ultracold Atoms, Massachusetts Institute of Technology, Cambridge, Massachusetts 02139, USA}
\affiliation{Research Laboratory of Electronics, Massachusetts Institute of Technology, Cambridge, Massachusetts 02139, USA}

\author{Shai Tsesses}
\thanks{These authors contributed equally}
\affiliation{MIT-Harvard Center for Ultracold Atoms, Massachusetts Institute of Technology, Cambridge, Massachusetts 02139, USA}
\affiliation{Research Laboratory of Electronics, Massachusetts Institute of Technology, Cambridge, Massachusetts 02139, USA}

\author{Peiran Niu}
\affiliation{MIT-Harvard Center for Ultracold Atoms, Massachusetts Institute of Technology, Cambridge, Massachusetts 02139, USA}
\affiliation{Research Laboratory of Electronics, Massachusetts Institute of Technology, Cambridge, Massachusetts 02139, USA}

  
 \author{Adrian Menssen}
\affiliation{MIT-Harvard Center for Ultracold Atoms, Massachusetts Institute of Technology, Cambridge, Massachusetts 02139, USA}
\affiliation{Research Laboratory of Electronics, Massachusetts Institute of Technology, Cambridge, Massachusetts 02139, USA}

\author{Wenchao Xu}
\affiliation{Institute for Quantum Electronics, Department of Physics, ETH Z\"{u}rich, Z\"{u}rich 8093, Switzerland}
\affiliation{Laboratory for Nano and Quantum Technologies, Paul Scherrer Institut, CH-5232 Villigen PSI, Switzerland}  

\author{Valentin Walther}
\affiliation{Department of Chemistry, Purdue University, West Lafayette, Indiana 47907, USA}
\affiliation{Department of Physics and Astronomy, Purdue University, West Lafayette, Indiana 47907, USA}

\author{Uro\v{s} Deli\'{c}}
\affiliation{Vienna Center for Quantum Science and Technology, Atominstitut, TU Wien, Stadionallee 2, 1020 Vienna, Austria}
\affiliation{University of Vienna, Faculty of Physics, Boltzmanngasse 5, A-1090 Vienna, Austria}

\author{Soonwon Choi}
\affiliation{MIT-Harvard Center for Ultracold Atoms, Massachusetts Institute of Technology, Cambridge, Massachusetts 02139, USA}
\affiliation{Center for Theoretical Physics — a Leinweber Institute,
Massachusetts Institute of Technology, Massachusetts 02139, USA}

\author{Mikhail   D. Lukin}
\affiliation{MIT-Harvard Center for Ultracold Atoms, Massachusetts Institute of Technology, Cambridge, Massachusetts 02139, USA}
\affiliation{Department of Physics, Harvard University, Cambridge, Massachusetts 02138, USA}

\author{Vladan Vuleti\'{c}}
\affiliation{MIT-Harvard Center for Ultracold Atoms, Massachusetts Institute of Technology, Cambridge, Massachusetts 02139, USA}
\affiliation{Research Laboratory of Electronics, Massachusetts Institute of Technology, Cambridge, Massachusetts 02139, USA}

\begin{abstract}
Rydberg atoms represent a platform underpinning many recent developments in quantum computation, simulation, sensing, and metrology. They further facilitate optical nonlinearity at the single-photon level when coupled to photons propagating in atomic clouds, which form collective atomic excitations called Rydberg polaritons, strongly interacting with each other. Here, we experimentally explore interactions between a Rydberg polariton in an atomic ensemble and a single, adjacent, Rydberg atom. We discover three different regimes of quantum dynamics corresponding to polariton blockade, coherent exchange, and probabilistic hopping, which are defined by their distinct transmission characteristics, with a transition through an exceptional point occurring between blockade and coherent exchange. We investigate the applications of such interactions for fast, non-destructive detection of Rydberg atoms and present proof-of-principle demonstrations for their potential application in nonlinear photonic networks.

\end{abstract}

\maketitle

\subsection{Introduction} 

Interactions between atoms in high-lying energy states,
also known as Rydberg atoms~\cite{RydbergAtom_generalpaper}, are playing a crucial role in many recent discoveries and technological advancements.
Manifesting in dipolar~\cite{de2019observation} or van der Waals~\cite{bernien2017probing} mechanisms, Rydberg interactions have
enabled the generation of squeezed atomic states ~\cite{spinsqueezing_antoine, spinsqueezing_kaufman, spinsqueezing_monika} for quantum metrology, the development of neutral atom quantum processors with high-fidelity gates ~\cite{evered_high_fidelity, thompson_highfidelity, manuel_highfidelity}, the quantum simulation of complex many-body phases~\cite{sepehr_quantumphases,scholl2021quantum,giulia_spinliquids} and strong optical nonlinearity at the single-photon level ~\cite{peyronel2012quantum, Srakaew2023, drori2023quantum}.  

Specifically, when photons propagating in atomic ensembles are coupled to Rydberg atoms via electromagnetically-induced transparency (EIT) ~\cite{misha_dark_state_polariton}, they hybridize into collective excitations known as Rydberg polaritons, which can interact strongly with other Rydberg polaritons or nearby Rydberg atoms. Past studies have shown that a Rydberg polariton colliding with a stationary Rydberg atom undergoes increased absorption~\cite{peyronel2012quantum,ourRydbergDetection} or a phase shift~\cite{thompson_symmetry_protected, vaneecloo_cavity_eit}, with applications such as single-photon transistors and switches~\cite{baur2014single,gorniaczyk2014single,tiarks2014single} and Rydberg atom imaging~\cite{gunter2012interaction,gunter_transport}. While the interaction between individual Rydberg atoms and polaritons has the potential to enable interesting applications such as nonlinear quantum metasurfaces built from atom arrays~\cite{bekenstein2020quantum} and 
photonic quantum gates via a nonlinear photonic network made up of atomic ensembles~\cite{PhysRevLett.123.113605}, such interactions have not been widely explored to date~\cite{thompson_symmetry_protected, PhysRevLett.123.113605, PhysRevApplied.19.014017}.

Here, we explore the full range of dipolar interactions between propagating Rydberg polaritons and an adjacent Rydberg atom. Our system consists of a single Rydberg atom, prepared inside an atomic ensemble through the self-blockade mechanism~\cite{lukin_blockade, ourRydbergDetection}, and separately trapped neighboring detection ensembles. Atoms in the detection ensembles are optically coupled to a Rydberg state under the EIT condition, where incoming photons are converted into Rydberg polaritons. We detect the presence of the Rydberg atom and the influence of dipolar interaction on the polaritons via the polaritons' transmission through the ensemble, while tuning the atom-polariton interaction strength through the inter-ensemble distance. We observe three distinct interaction regimes with different transmission characteristics: polariton blockade, coherent atom-polariton exchange, and probabilistic hopping. At short distances, the Rydberg atom energetically blocks polariton formation, leading to increased polariton absorption (blockade). At intermediate distances, the dipolar interaction produces coherent exchange of atomic excitations, resulting in non-monotonous transmission. At long distances, the interaction becomes perturbative, resulting in probabilistic hopping of the Rydberg excitation to neighboring ensembles.

\begin{figure}[htbp]
\raggedright
\includegraphics[width=0.48\textwidth]{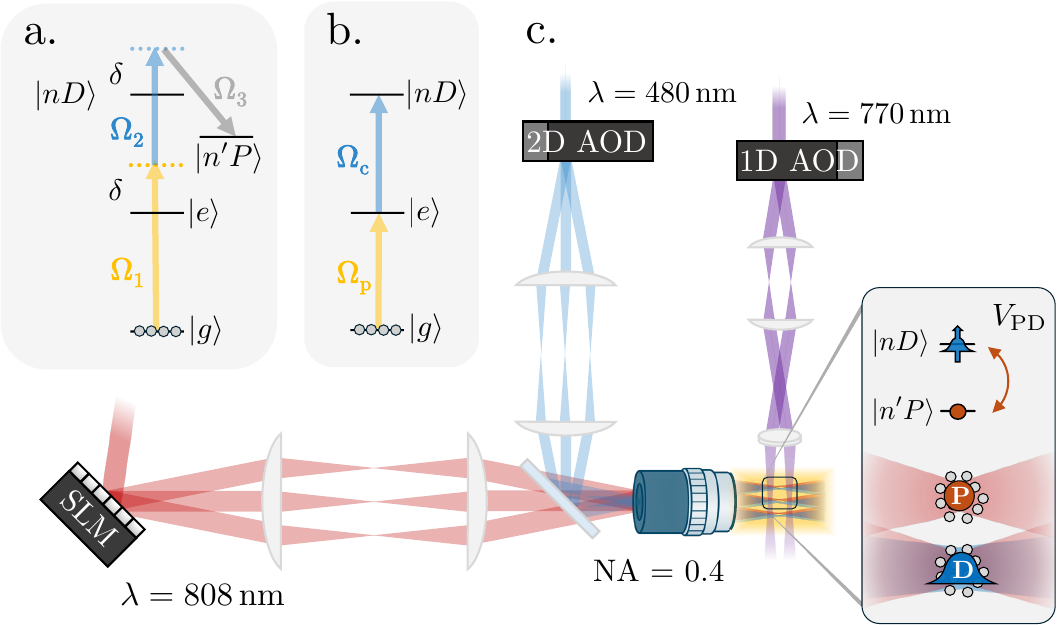}
\caption{\textbf{Experimental setup and level scheme: a.} Three-photon scheme for preparing a single Rydberg atom in the state $\ket{n'P} = \ket{75\textrm{P}_{3/2}, m_J = 3/2}$. The first and third legs of the three-photon transition ($780$\,nm and microwave fields with Rabi frequencies $\Omega_{1}$, $\Omega_{3}$) are detuned by $\delta = +100\,$MHz from resonance of their respective transitions, while the second leg ($480$\,nm, $\Omega_{2}$) is on-resonance.
\textbf{b.} Rydberg EIT scheme couples ground-state atoms to the Rydberg state $\ket{nD} =\ket{74\textrm{D}_{5/2}, m_J = 5/2}$. Both the probe and control fields ($\Omega_{p}$, $\Omega_{c}$) are on-resonance with their respective transitions. \textbf{c.} Schematic of the setup illustrating the trap configuration, which loads small atomic ensembles using a combination of red-detuned $808\,\textrm{nm}$ SLM traps (red) and blue-detuned $770\,\textrm{nm}$ AOD sheets (purple). A site-selective $480\,\textrm{nm}$ control field array (blue) co-propagates with the SLM trap light, while the global $780$\,nm probe field counter-propagates (yellow). Inset: Dipole-dipole interactions $\textrm{V}_\textrm{PD}$ between a stationary Rydberg atom in $\ket{n'P}$ and traveling Rydberg polaritons in $\ket{nD}$ lead to modulation of probe EIT transmission.}
\label{figure1}
\end{figure}

Interestingly, we identify a transition through an exceptional point~\cite{heiss2012physics} -- the coalescence of energy eigenvalues in a non-Hermitian system -- between the polariton blockade and coherent exchange regimes, reminiscent of the dissipative phase transition in an open central spin system~\cite{kessler2012dissipative}. To showcase the applicability of Rydberg atom-polariton interaction, we employ it for fast, remote, and non-destructive detection of Rydberg atoms, exemplified by observing coherent atomic oscillations via repeated polariton transmission measurements. Finally, we perform proof-of-principle demonstrations of the building blocks required to construct multi-rail photonic networks with state memory~\cite{PhysRevLett.123.113605}. Our results elucidate the physics of dipolar interactions between Rydberg atoms and polaritons, as well as pave the way toward ensemble-based readout in neutral-atom quantum processors~\cite{zhang2025dual} and nonlinear photonic networks for quantum random walks of photons~\cite{random_walk}.

\subsection{Experimental setup}
Our experimental setup is illustrated in Fig. \ref{figure1}c. We load ensembles of $^{87}$Rb atoms~\cite{our_bec_paper} in an array of dipole traps ($\lambda = 808\,$nm, $w_0 = 4.9\,\mu$m) generated by a spatial light modulator (SLM) and a microscope objective (numerical aperture $\textrm{NA}=0.4$). After loading, we dynamically compress the ensembles (see Supplementary Movie 1) to reduce the axial cloud size and improve confinement with two sheets of blue-detuned beams (wavelength $\lambda = 770\,$nm, waists $w_{0,\{x, y\}} = \{4, 32\}\,\mu$m), generated by a one-dimensional acousto-optic deflector (AOD). After these steps, each dipole trap hosts an ensemble of $\sim 200$ atoms, with typical peak optical densities of $\textrm{OD}\approx{2}$ and root-mean-square sizes $\sigma_{\{x, y, z\}} = \{2, 2.5, 4\}\,\mu$m, far smaller than the minimal distance between traps in this work ($\gt 10 \,\mu$m). 

In the experiments presented below, we load 2--5 atomic ensembles, and detect the polaritons going through them via absorption imaging. Fig. \ref{figure1}b illustrates our Rydberg EIT configuration: a weak, $780\,\textrm{nm}$ EIT probe field (Rabi frequency $\Omega_{p}/(2\pi) \approx 100-300\,$kHz) addresses the transition between the ground state $\ket{g} = \ket{5S_{1/2}, F=2, m_F = 2}$ and the excited state $\ket{e} = \ket{5P_{3/2}, F=3, m_F = 3}$, while a strong, site-selective $480\,\textrm{nm}$ EIT control field ($\Omega_{c}/(2\pi) = 5-23\,$MHz) couples $\ket{e}$ to the Rydberg state $\ket{nD} =\ket{74\textrm{D}_{5/2}, m_J = 5/2}$. This is performed at a magnetic field of 9\,G, so as to spectrally resolve the desired Rydberg $m_J$ sublevel. 

If the probe and control fields satisfy the bare two-photon resonance (while $\Omega_p \ll \Omega_c$), the EIT condition is fulfilled, resulting in high probe transmission. In this work, the probe intensity is maximized while ensuring at most a single polariton is present in each ensemble at a given time, to avoid the self-blockade effect~\cite{peyronel2012quantum, dipolar_dephasing_hofferberth}. In each experiment, a single ensemble ("preparation ensemble") is used to prepare a single Rydberg atom in the state $\ket{n'P} = \ket{75\textrm{P}_{3/2}, m_J = 3/2}$ via a three-photon transition and the Rydberg blockade mechanism~\cite{ourRydbergDetection}, as illustrated in Fig.~\ref{figure1}a. The presence of a stationary Rydberg atom in $\ket{n'P}$ shifts the states $\ket{nD}$ within the nearby ensembles ("detection ensembles"), thereby inhibiting polariton formation and leading to probe absorption. Further information about our apparatus can be found in the Supplementary Material (SM)~\cite{SM}.

\subsection{Rydberg atom-polariton interaction regimes}

The stationary Rydberg atom in $\ket{n'P}$ and the $\ket{nD}$ Rydberg polaritons interact via dipolar interaction, with an energy splitting of the form $V_\textrm{PD}(r) = 2C_3 / r^3$, where $r$ is the atom-polariton distance and $C_3$ is the interaction coefficient. Atom-pair interaction simulations~\cite{Weber2017} indicate that the interactions between the two states chosen in this work are relatively isolated from other energy levels; therefore, the dynamics remains predominantly within the quantum states considered above~\cite{SM}. We classify different interaction regimes via absorption in a detection ensemble and extract the \textit{detection contrast}, defined as the ratio of photons lost to the number of detected photons in the absence of the Rydberg atom (bare EIT transmission), which reflects the local $\ket{n'P}$ population.

\begin{figure}[htbp]
\centering
\includegraphics[width=0.4\textwidth]{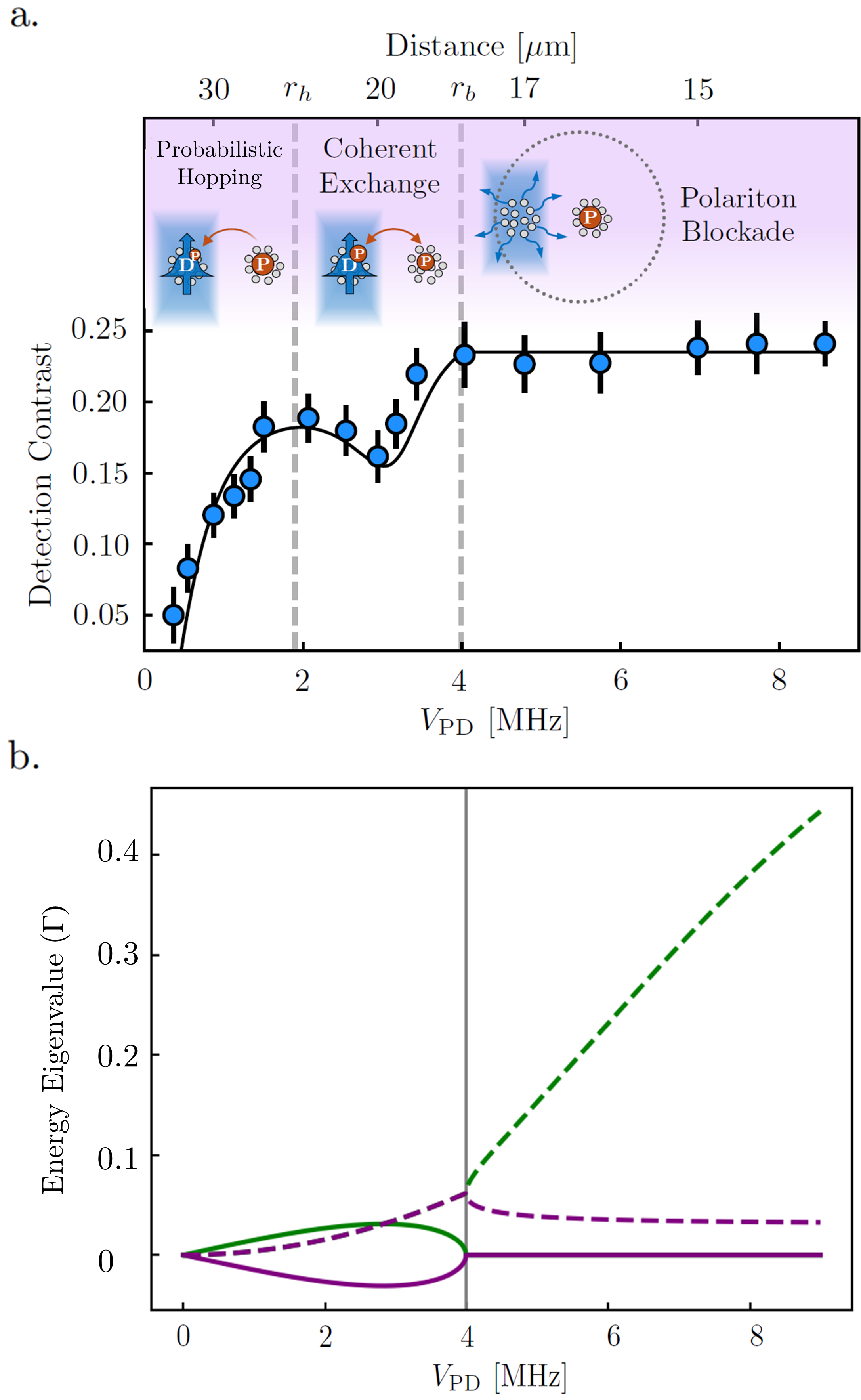}
\caption{\textbf{Classifying Rydberg atom-polariton interaction regimes: a.} Detection contrast (see main text) of probe photon transmission through the detection ensemble (blue), for a single Rydberg atom prepared in $\ket{n'P}$ within the nearby preparation ensemble (orange), as a function of the dipolar interaction strength V$_\textrm{PD}$ and inter-ensemble distance. After preparation of the Rydberg atom, the detection ensemble is probed for 10~$\mu$s, much longer than the polariton lifetime $\tau_d \approx  60\,\textrm{ns}$. Three distinct regimes, corresponding to probabilistic hopping, coherent exchange, and polariton blockade, can be identified. The vertical lines represent the interaction strengths at the hopping radius $r_h\approx 23~\mu$m and the blockade radius $r_b\approx 18~\mu$m for the EIT control Rabi frequency of $\Omega_{c}/(2\pi) =5.5\, \textrm{MHz}$, which was chosen to allow observation of all three regimes. A phenomenological theory curve (solid black line; see SM~\cite{SM}) closely matches the observed data.
\textbf{b.} Ab-initio calculation of two polariton energy eigenvalues of the Hamiltonian describing the system (see Eq. \ref{Eq:compact} in SM~\cite{SM}). We plot the real (solid) and imaginary (dashed) parts of the eigenvalues (in units of the intermediate state linewidth $\Gamma$) as a function of the dipole-dipole interaction strength, $V_\textrm{PD}$. An exceptional point clearly emerges at the interaction strength corresponding to the blockade radius (gray line). At high interaction strengths (right-hand side of the gray line), the polariton eigenvalues are imaginary (dissipative). At low interaction strengths (left-hand side of the gray line), the polariton eigenvalues have the same imaginary part (i.e., dissipation still occurs), but have a non-degenerate real part, with the same magnitude and an opposite sign (signifying the dressing of the polariton state due to the interaction with the atom).
}
\label{figure2}
\end{figure}

Our results are summarized in Fig.~\ref{figure2}a. The first regime, polariton blockade, occurs within the blockade radius $r \lt r_b$, which is defined as $V_\textrm{PD}(r_b) = \gamma_{\textrm{EIT}}$~\cite{PhysRevLett.123.113605, thompson_symmetry_protected, fleischhauer2005electromagnetically}, where $\gamma_{\textrm{EIT}}$ is the EIT linewidth. The strong interactions in this regime prevent polariton formation, correspondingly leading to a high detection contrast in Fig.~\ref{figure2}a, irrespective of $V_\textrm{PD}$.

As the atom-polariton distance increases, the second regime emerges, in which the dipolar interaction strength is sufficiently small to allow polariton formation. Coherent exchange of Rydberg excitations $\ket{n'P} \leftrightarrow \ket{nD}$ commences between the atom and the polariton, giving rise to a non-monotonic behavior of the detection contrast in Fig.~\ref{figure2}a. Minimal contrast is observed at a distance for which a full exchange $\ket{n'P} \rightarrow \ket{nD} \rightarrow \ket{n'P}$ is completed within the characteristic time for the polariton to leave the ensemble, known as the polariton lifetime $\tau_d=  \textrm{OD}\, \frac{\Gamma}{\Omega_c^2}$, with $\Gamma$ being the linewidth of the $\ket{5P_{3/2}}$ state. At this distance, ideally, a polariton would exit the detection ensemble unperturbed and lead to zero detection contrast; however, in our system, the visibility of this effect is limited by the lower value of OD, positional dephasing, and additional decay and decoherence mechanisms. The regime of coherent exchange occurs for intermediate distances $r_b \lt r \lt r_h$, where the hopping radius $r_h$ is defined as the distance at which the interaction timescale equals the polariton lifetime ($1 / V_\textrm{PD}(r_h) = \tau_d $). The hopping radius is related to the blockade radius as $r_h = \sqrt{\textrm{OD}_b} \,r_b$~\cite{PhysRevLett.123.113605}, where $\textrm{OD}_b$ is the optical density per blockade radius. In our case, $\textrm{OD}_b = \textrm{OD}$ since the ensemble is smaller than the blockade radius, enforcing $r_h \gt r_b$. 

Finally, at distances beyond the hopping radius, $r \gt r_h$, the interaction is weak and the detection contrast in Fig. \ref{figure2}a drops quadratically with $V_\textrm{PD}$, consistent with rare and probabilistic hopping of the $\ket{n'P}$ excitation. The full range of observed data matches our phenomenological stochastic hopping theory (see SM~\cite{SM}) which captures all key features of our experiment.

\subsection{Exceptional point in the interaction of a Rydberg atom and a Rydberg polariton}

It is clear from Fig.~\ref{figure2}a that the detection contrast quantitatively identifies the transitions between the different interaction regimes. It is worth noting that switching between interaction regimes may cause an abrupt change to the steady-state properties of the system, which is often the sign of a phase transition in a many-body system such as ours. Specifically, the interaction in the blockade regime introduces significant dissipation to incoming probe photons, whereas in the exchange regime, polaritons can form and oscillations arise in the effective atom-polariton system. Thus, in contrast to most non-Hermitian systems, in our system the coherent dynamics resurface when the strength of the coherent interaction is \textit{decreased}.

To investigate this fundamental change in behavior, we calculate in Fig.~\ref{figure2}b the polariton eigenvalues of the Hamiltonian describing a single excitation, following the established formalism~\cite{bienias2014scattering,thompson_symmetry_protected}. From ab-initio considerations alone~\cite{SM}, we find a clear shift occurring exactly at the interaction strength consistent with interface between the blockade and exchange regimes. The eigenvalues are non-degenerate and purely imaginary in the blockade regime, yet they become complex in the exchange regime, acquiring a degenerate imaginary part and a non-degenerate real part. This result verifies that our system undergoes, by definition, a transition through an exceptional point~\cite{heiss2012physics}. 

Exceptional points are degeneracies in non-Hermitian systems, where two or more energy eigenvalues coalesce, creating a system highly sensitive to small external perturbations that can exhibit enhanced sensing capabilities~\cite{hodaei2017enhanced,chen2017exceptional,kononchuk2022exceptional}. In many-body systems, a transition through an exceptional point also routinely signifies a dissipative phase transition ~\cite{Fitzpatrick2017Observation,fink2018signatures,ferioli2023non,kuzmin2024observation}, arising from the interplay of coherent interaction and coupling to the environment. Our system bears a particular resemblance to the open central spin model (see~\cite{SM}), which indeed exhibits such a phase transition~\cite{kessler2012dissipative}. However, even though the Hamiltonian description is the same, and the transition through the exceptional point occurs with the same order parameter, there are meaningful differences. The most important difference is the hard-core boson nature of Rydberg excitations, which does not exist in the central spin model and limits our many-body system to effective two-body dynamics.

\begin{figure}[htbp]
\centering
\includegraphics[width=0.5\textwidth]{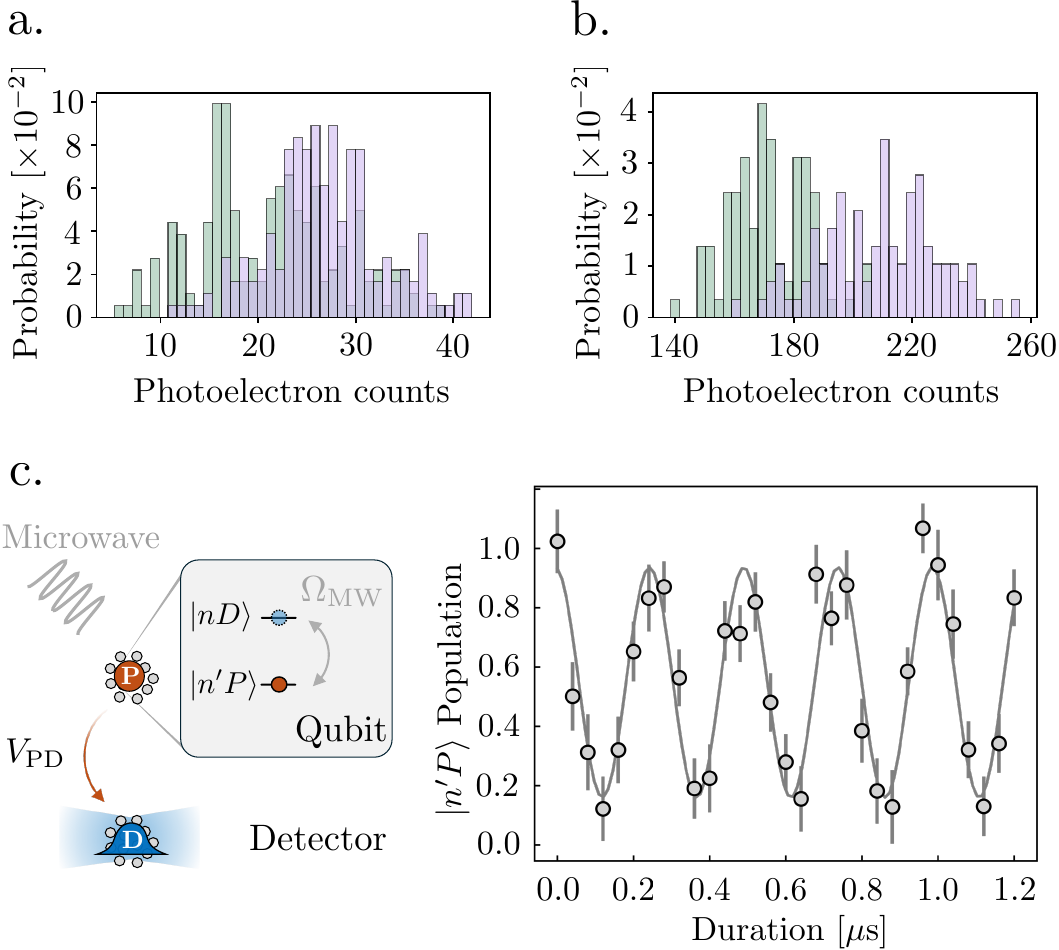}
\caption{\textbf{Rydberg EIT as a remote detector of single Rydberg atoms: a.} Single-shot histogram for 30\,$\mu$s of readout, showing the difference in detected photons exiting the detector ensemble with (green) and without (purple) a Rydberg atom nearby. The measurement fidelity is $47(5)\,\%$. \textbf{b.} Using repeated preparation and detection (20 repetitions of $10\,\mu$s detection), we achieve a fidelity of $79(4)\,\%$. \textbf{c.} Following a global microwave drive, state-selective and remote detection of the Rydberg qubit is performed, showing long-lived coherent microwave Rabi oscillations.}
\label{figure4}
\end{figure}

\subsection{Remote and repeated Rydberg qubit readout}

After identifying the interaction regimes in dipolar Rydberg atom-polariton interactions, we turn to demonstrating their potential applications. A direct application is fast, non-destructive, state-selective, and \textit{remote} readout of a single Rydberg qubit. Until now, this task was only possible if the atom was placed within the detection ensemble~\cite{ourRydbergDetection,du2025imaging}. For an efficient fast detector, high detection contrast and probe rate are desirable, yet are difficult to achieve simultaneously. A strong control field increases the possible probe rate but reduces the blockade radius ($r_b \propto \Omega_c^{-2/3}$), limiting access to the blockade regime where the detection contrast is the highest. The blockade regime also preserves the state of the Rydberg atom, and allows multiplexing the signal from multiple detector ensembles to improve readout fidelity, while the inherent non-destructive nature of our detection method minimizes atom loss. 

Due to our large dipole trap waist ($w_0 = 4.9\,\mu$m) and relatively small NA=0.4 of the microscope objective, the largest signal-to-noise ratio for detection was achieved in the probabilistic hopping regime, rather than the blockade regime. The detector parameters were manually optimized to a high control field ($\Omega_{c}/(2\pi) = 20\,\textrm{MHz}$), an average optical density of $\textrm{OD}\approx{1.5}$, $\approx 200$ atoms per ensemble, and a qubit-detector separation of $20\,\mu$m (more details in SM~\cite{SM}). 

Fig.~\ref{figure4}a shows the results for single-shot detection of a Rydberg atom, with a combined preparation and detection fidelity of $47(5)\,\%$ to distinguish between qubit states within $30\,\mu$s of readout. This is the probability of correctly preparing and identifying the state of the Rydberg atom (i.e., a random guess has fidelity of $0\,\%$). Since the detection scheme leads to minimal atom loss in the ensemble, repeated readout sequences are possible. By performing this process for 20 repetitions of $10\,\mu$s detection time, and with the same ensemble, we achieved a fidelity of $79(4)\,\%$ as shown in Fig.~\ref{figure4}b, still within a timescale far smaller than current prevalent methods~\cite{zeiher_continuous,chiu2025continuous,norcia2024iterative}. 

\begin{figure*}[htbp]
\centering
\includegraphics[width=0.8\textwidth,trim= 0 0 200 0,clip]{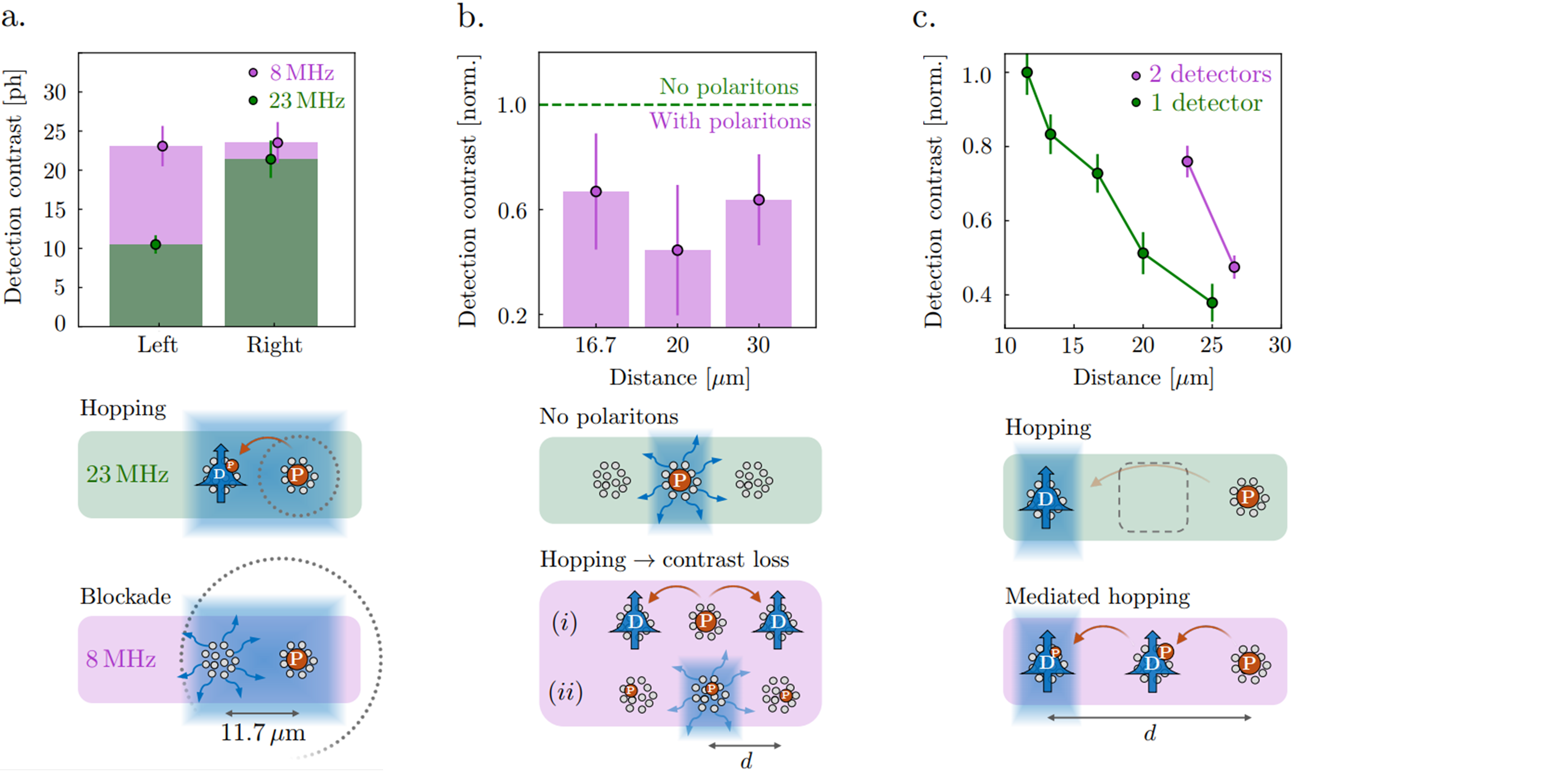}
\caption{\textbf{Hopping and blockade in a nonlinear photonic network: a.}  
Detection contrast (in photons) in both ensembles following preparation of a Rydberg atom in the right ensemble, with $11.7\,\mu\textrm{m}$ of inter-ensemble distance ($V_\textrm{PD}/h = 14.8\,\textrm{MHz}$). Simultaneous readout of both ensembles is performed for two EIT control fields, $\Omega_c/(2\pi) = 23\,\textrm{MHz}$ ($r_b \approx 9\,\mu$m; green) and  $\Omega_c/(2\pi) = 8\,\textrm{MHz}$ ($r_b \approx 13~\mu$m; purple). When the distance between ensembles is smaller than the blockade radius (lower purple panel), the probe absorption is the same for both ensembles, which indicates that the left ensemble is blocked by the presence of a Rydberg atom in the right ensemble. On the other hand, when the distance is larger than the blockade radius (upper green panel), the probe absorption is different, and polaritons are allowed to propagate in the left ensemble.
\textbf{b.} Detection contrast in the middle ensemble, which reflects the local $\ket{n'P}$ Rydberg population. The initial population (green) decreases due to the presence of polaritons in the neighboring detection ensembles (purple), which indicates exchange between the prepared Rydberg atom and neighboring Rydberg polaritons. Results are normalized according to the initial $\ket{n'P}$ Rydberg population. Note that the large control field used for these measurements ($\Omega_{c}/(2\pi) \approx 20\, \textrm{MHz}$) places the blockade radius at $r_b \approx 9\,\mu$m and the hopping radius at $r_h \lt 14\,\mu$m.
\textbf{c.} Detection contrast (normalized) in the leftmost ensemble as a function of distance (green) increases in the presence (purple) of an intermediate ensemble, which mediates exchange between the Rydberg atom in the rightmost ensemble to the leftmost ensemble. The blue shaded regions denote ensembles where we measure and plot the absorption of probe EIT light, which is proportional to the Rydberg population within the ensemble. Other than the geometry, experimental parameters are the same as in \textbf{b.}  
}
\label{figure3}
\end{figure*}  

Using our repeated readout scheme, we demonstrate state-selective, remote detection of a single Rydberg qubit. Following the preparation of a single atom in the $\ket{n'P}$ state, a global $\nu = 5.214\, \textrm{GHz}$ microwave field is applied, which drives Rabi oscillations between the $\ket{n'P}$ and $\ket{nD}$ states at a frequency of $\Omega_\textrm{MW}/(2\pi) = 4.04(4)\, \textrm{MHz}$. Meanwhile, the microwave field has no effect on the ground-state atoms in the detector ensemble. After a variable microwave duration, we probe the detector ensemble with Rydberg polaritons, as shown in Fig.~\ref{figure4}c, ensuring that it measures the qubit state even at a large distance of $20\,\mu$m. The fidelity of our repeated detection, accounting for both preparation and detection errors, is comparable to the previously reported single-shot fidelity in other, non-scalable detector geometries~\cite{ourRydbergDetection}.

\subsection{Towards nonlinear photonic networks with atomic ensemble arrays} 

The presence of a Rydberg atom inside an atomic ensemble has been used to control the amplitude and phase of transmitted photons~\cite{thompson_symmetry_protected,ourRydbergDetection}. Our system, however, opens the opportunity for properly placed Rydberg atoms to control the flow of photons through an array of ensembles, producing a versatile nonlinear photonic network acting differently according to the atom-polariton interaction regime, with applications such as photonic quantum gates~\cite{PhysRevLett.123.113605}, components for optical quantum networks~\cite{PhysRevA.81.052329, PhysRevLett.128.060502, PhysRevLett.132.053001} or optical neural network implementations~\cite{shen2017deep,wang2023image}. The building blocks of such a nonlinear photonic network are switches, which determine whether or not photons are transmitted; state memory, whereby the network configuration changes given previous operations; and connectivity, which determines functionality and capabilities. In the following, we explore these building blocks within our system, choosing the appropriate interaction regime for each. 

In Fig.~\ref{figure3}a, we demonstrate a new mechanism for an all-optical Rydberg switch~\cite{baur2014single,gorniaczyk2014single,tiarks2014single}, based on the blockade radius dependence on the EIT control field ($r_b \propto \Omega_c^{-2/3}$). We prepare a Rydberg atom in the ensemble on the right, and detect the transmission simultaneously on both ensembles. When the ensemble on the left is outside the blockade radius $r_b$, there is a significant disparity in the detection contrast, and the polaritons on the left are allowed to propagate. On the contrary, when the ensemble on the left is within the blockade radius, we find that detection contrast in both ensembles is similar, i.e., polariton formation is prevented in both. We note that this all-optical control mechanism was possible solely due to the non-local nature of the interaction. 

Intriguingly, outside the blockade regime, the detection contrast constitutes a nonlinear property, requiring more than one photon, which is unusual in EIT physics in the absence of polariton self-interaction~\cite{gorshkov2011photon}. A polariton must first exchange its state with an atom, creating a stationary $\ket{n'P}$ state, for the transmission of subsequent photons to decrease. In this manner, our system can exhibit a memory of previous interaction events~\cite{distante2017storing} in the coherent exchange and probabilistic hopping regimes, while supporting several interaction events, which are a prerequisite for network connectivity and fan-out. 

To test these capabilities, we produce an array of three ensembles, prepare a single Rydberg atom in the $\ket{n'P}$ state in one of the ensembles, and monitor its location within the array. In Fig.~\ref{figure3}b, we prepare a Rydberg atom in the middle ensemble, and generate polaritons in two neighboring ensembles. For various inter-ensemble distances (all within the probabilistic hopping regime), it is evident that the $\ket{n'P}$ population is reduced by the presence of nearby polaritons (lower purple panel) compared to the initial population (upper green panel;~\cite{SM}): Part of the $\ket{n'P}$ population was transferred to the other ensembles. 

In Fig.~\ref{figure3}c, we prepare a Rydberg atom in the rightmost ensemble and measure the transmission of the leftmost ensemble as a function of distance. The detection contrast increases when an intermediate ensemble is present (lower purple panel) compared to when it is absent (upper green panel), indicating that the $\ket{n'P}$ excitation transferred to the leftmost ensemble both directly and through the intermediate ensemble. Thus, full connectivity of the 3-ensemble network is confirmed, while also validating the working principle of a proposed photonic quantum gate~\cite{PhysRevLett.123.113605}. We note that in this experiment, the intermediate ensemble is within the hopping radius $r_h$ of its neighbors (i.e., in the coherent exchange regime), while the bare interaction between the rightmost and leftmost ensembles is 8 times weaker, placing them well within the probabilistic hopping regime~\cite{SM}.     

\subsection{Conclusion and outlook}

In summary, we have observed the full range of dipolar interactions between a Rydberg polariton in an atomic ensemble and a single, nearby, Rydberg atom. The three possible interaction regimes -- polariton blockade, coherent exchange, and probabilistic hopping -- each have distinct transmission characteristics, with a transition through an exceptional point occurring between blockade and coherent exchange. We further investigated fast, remote and non-destructive detection of Rydberg atoms, and explored the building blocks of nonlinear photonic networks.

Future studies may utilize our work as a convenient infrastructure for implementing photonic quantum gates~\cite{baur2014single,gorniaczyk2014single,tiarks2014single}, since our results imply that the proposed mechanism for a controlled-Z gate is feasible~\cite{PhysRevLett.123.113605}. The controlled phase of such a gate could be tuned continuously depending on the strength of the exchange coupling, which can be engineered via the atom-polariton separation. Furthermore, by increasing the number of ensembles and changing both the array geometry and the EIT control illumination scheme, it will become possible to probe quantum walks of photons~\cite{random_walk,gong2021quantum} in a new system with strong photon-photon interactions and state memory. In such a system, the number and position of stationary Rydberg atoms, propagating Rydberg polaritons, and the polariton lifetime, will all be controllable, resulting in a rich testbed for coherent and incoherent many-body physics. 

Although the readout fidelity achieved in this work is not yet sufficient for immediate implementation in state-of-the-art quantum-information-related applications, we outline a path to improve it to the required level of $99\,\%$ ~\cite{SM}. One possible improvement we already attempted is using several detector ensembles per atom -- this approach is beneficial in the blockade regime, yet it provides no advantage in the coherent exchange and probabilistic hopping regimes, as the total Rydberg population is conserved, along with the total detection contrast (see~\cite{SM}). With improved readout fidelity, our scheme can be scaled to interface arrays of ensemble detectors with single-atom arrays in neutral atom quantum processors~\cite{zhang2025dual}. Such a platform could facilitate non-destructive detection in free space within tens of $\mu$s or less -- orders-of-magnitude faster than currently used methods -- while being compatible with continuous array operation at a constant finite magnetic field, as well as with existing architectures for neutral-atom quantum processors~\cite{dolev_logicalcomputer,66s8-jj18}. 

\subsection{Acknowledgements}
We would like to thank Alexey Gorshkov, as well as Aashish Clerk and his group for insightful discussions. We would also like to thank Zachary Vendeiro and Enrique Mendez for support with \textsf{M-LOOP}, and Matthew Peters for support with our imaging camera. E.H.Q. acknowledges the support of the Natural Sciences and Engineering Research Council of Canada (NSERC). S.T. acknowledges support from the Rothschild fellowship of the Yad Hanadiv foundation, the VATAT-Quantum fellowship of the Israel Council for Higher Education, the Helen Diller Quantum Center postdoctoral fellowship and the Viterbi fellowship of the Technion – Israel Institute of Technology. VW acknowledges support from the National Science Foundation (NSF) under Award $\#$PHYS-2409630. This material is based upon work supported by the U.S. Department of Energy, Office of Science, National Quantum Information Science Research Centers, Quantum Systems Accelerator. Additional support is acknowledged from the NSF-funded Center for Ultracold Atoms (grant number PHY-2317134), the DARPA ONISQ program (grant number 134371-5113608), the IARPA ELQ program (grant number W911NF2320219), and the NSF QLCI Q-SEnSE (grant number QLCI-2016244).

\bibliography{reference}

\clearpage

\beginsupplement

\section{Supplemental Material}

\subsection{Experimental Setup}
Our machine-learning-enhanced ensemble loading procedure is previously described in ref.~\cite{our_bec_paper}.
The trapping light ($\lambda = 808\,$nm) generated by the SLM (Hamamatsu x13138-02) is focused through our commercial microscope objective (Mitutoyo M Plan Apo NIR B 20X 378--867--5, $\textrm{NA}=0.4$) to a $1/e^2$ beam waist of $w_0 = 4.9\,\mu$m. 

After loading, each trap contains $\approx 1500$ atoms.
To make the atomic clouds more round, we compress the ensembles axially in 80~ms with two sheets of blue-detuned beams ($\lambda = 770\,$nm) generated with a one-dimensional AOD (DTSX-400-800.850).
After this procedure, each trap contains $\approx 200$ atoms with rms sizes $\sigma_{\{x, y, z\}} = \{2, 2.5, 4\}\,\mu$m.


After ensemble compression, we increase the magnetic field to $9\,$G along the propagation axis of the tweezer light, which defines the quantization axis for the remainder of the sequence. We optically pump the atoms in $4\,$ms with $\sigma^+$ polarized light on the $D_2$-line $F=2 \rightarrow F'=2$ transition to the dark state $\ket{g} = \ket{5S_{1/2}, F=2, m_F = 2}$, while repumping with unpolarized light on the $F=1 \rightarrow F'=2$ transition.



\subsection{Three-photon Rydberg preparation}
We choose to work with principal quantum numbers $n=74,n'=75$, which are sufficiently large for Rydberg blockade during state preparation and detection, without significantly increasing the self-blockade effect of Rydberg polaritons\,\cite{peyronel2012quantum}. To prepare a single Rydberg excitation within an ensemble, we use a global $780\,$nm beam, a $5.214\,$GHz microwave (MW) source, and a $480\,$nm site-selective addressing beam ($w_0 = 2.3\,\mu$m), which is used to selectively illuminate either the preparation or the detection ensembles. Since the $480\,$nm light is on-resonance, we blue-detune both the $780\,$nm and MW frequencies by $\delta = 100\,$MHz to avoid populating the intermediate states during preparation.

Our preparation scheme and parameters used in this experiment are similar to those described in~\cite{ourRydbergDetection}. We prepare a single Rydberg atom in the state $\ket{75\textrm{P}_{3/2}, m_J = 3/2}$ with a van der Waals interaction coefficient $C_6 = 45\,$GHz$\,\mu$m$^6$, and we observe a power-broadened three-photon linewidth of $\gamma_{3\textrm{ph}}/(2\pi) \approx 500\,$kHz, corresponding to a blockade radius of $\left(C_6/\gamma_{3\textrm{ph}}\right)^{1/6} = 6.7\,\mu$m, which is sufficient to block the preparation of multiple Rydberg excitations in a single ensemble.
All three legs of the three-photon transition are turned on simultaneously, 
and we observe saturation of Rydberg population with a characteristic timescale of $\approx 3\,\mu$s in the overdamped regime. We do not observe collective oscillations consistent with the preparation of a $\ket{W}$ state, which we attribute to fast dephasing of the $\ket{W}$ state (a few hundred nanoseconds has been experimentally observed\,\cite{csadams_storage}). Therefore, we prepare a single Rydberg excitation randomly distributed within the cloud with an upper-bounded-fidelity of $N/N+1$, where $N$ is the number of atoms in the cloud\,\cite{buchler_rateequation_preparation}. In practice, our preparation fidelity was circa 90\%, consistent with previous studies~\cite{ourRydbergDetection}.

\subsection{Site-selective preparation and detection using an AOD on the control field}

Site-selective preparation and detection are performed using a $480 \textrm{nm}$ AOD on the $\sigma^+$ polarized control beam, which co-propagates with the trapping light (see Fig. 1c of the main text). The remaining fields used for preparation and detection are global beams addressing all traps. For EIT detection, the resonant $\sigma^+$ polarized $780\,$nm probe beam ($w_0 = 1\,$mm) counter-propagates along the tweezer axis. For three-photon preparation, the linearly polarized $780\,$nm beam ($w_0 = 1\,$mm) propagates along the transverse axis of traps. 
The microwave (MW) field used for preparation and Rabi driving is an unpolarized field generated by a stub-terminated SMA cable~\cite{PhysRevLett.119.053202}.

To selectively prepare a Rydberg atom in one ensemble and perform subsequent readout using a neighboring ensemble, we switch the $480 \,\textrm{nm}$ AOD frequency mid-sequence in $8\,\mu$s. This time duration is sufficiently small compared to the lifetimes of the Rydberg states ($\gtrsim 100 \mu$s), such that the Rydberg atom preparation time can be neglected. For all experiments presented in this work, the light is turned off before AOD switching with the use of an additional AOM to prevent ensembles from being illuminated by control light during the AOD response time. In our sequence, we perform state preparation, switch the AOD frequency, and wait $10\,\mu$s before EIT detection.

\subsection{Choice of Rydberg states: $D-P$ vs. $S-P$}

Typical Rydberg atom array experiments choose a principal quantum number $n$ and tweezer spacing such that the system operates in the van-der-Waals regime, where the Rydberg blockade is dominated by coupling to one isolated Rydberg state. However, within small mesoscopic ensembles, the mean-free-path of atoms can be on the order of $\sim 1\,\mu$m for densities of $\sim 10^{12}\,$cm$^{-3}$\,\cite{our_bec_paper}, in which resonant dipole-dipole interactions between many pair states can lead to complicated energy landscapes. 
Isolating a pair of Rydberg states to engineer clean dipolar interactions is challenging in atomic ensembles and requires careful choice of Rydberg states and external electric/magnetic bias fields~\cite{zeeman_degeneracy_saffman, browaeys_efield_tuning}. For example, interactions between Rydberg polaritons at short distances can lead to finite populations in other degenerate Zeeman sublevels, thus leading to a reduction in EIT transmission over the course of probing\,\cite{dipolar_dephasing_hofferberth}. 

\begin{figure}[htbp]
\centering
\includegraphics[width=0.40\textwidth,trim= 0 0 0 0,clip]{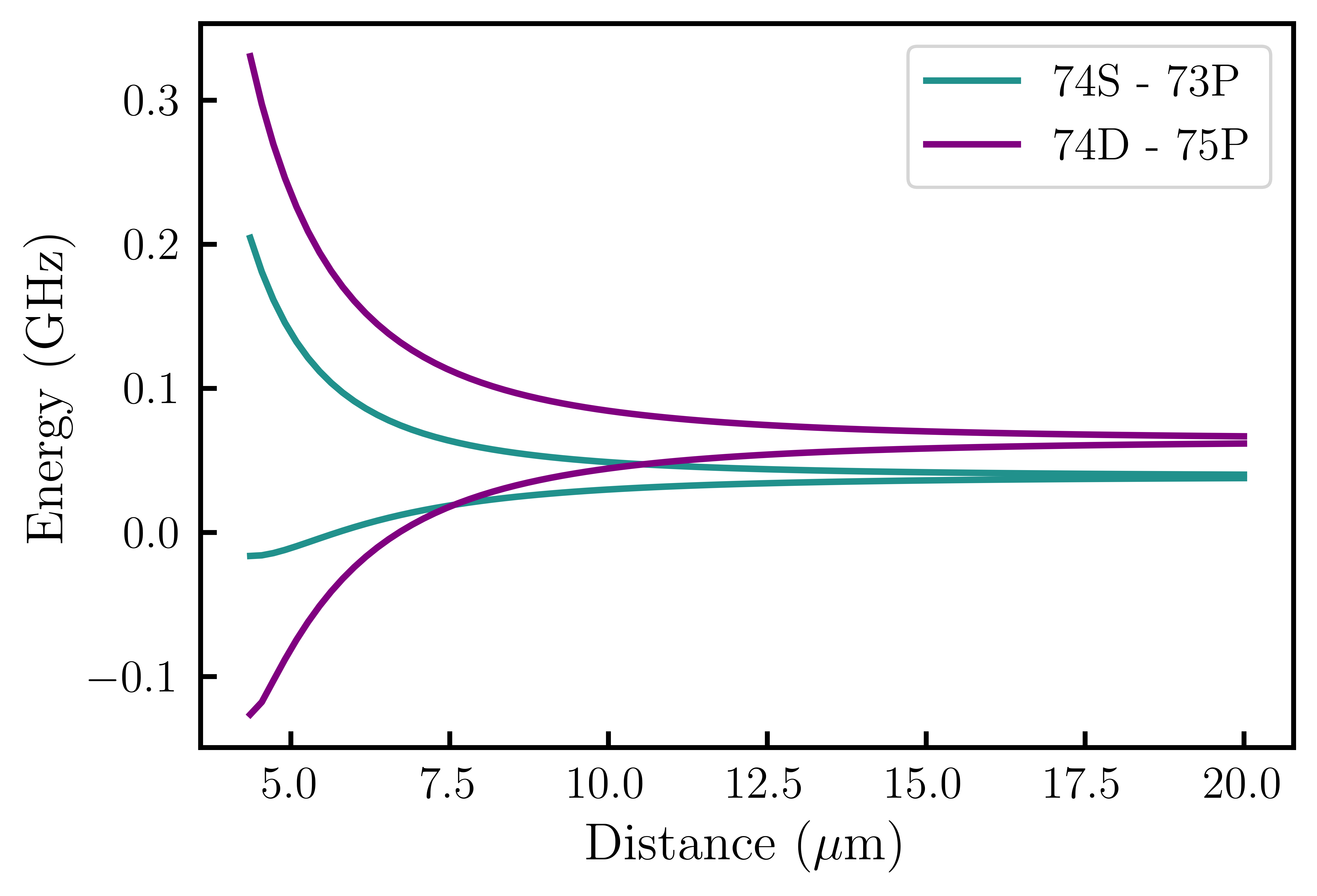}
\caption{\textbf{DP vs SP:} An example showcasing the purity of exchange interactions for $D-P$ states in comparison to  $S-P$ states (calculated with ``PairInteraction"~\cite{Weber2017}). The states are chosen so that both the D and S states share the same Rydberg $n$ level, while the P states in each case are selected to maximize interaction strength.
}
\label{DPvsSP}
\end{figure}

We choose to work with Rydberg polaritons in the state $\ket{74\textrm{D}_{5/2}, m_J = 5/2}$ with $C_6\approx-1600\,$GHz\,$\mu$m$^6$, while ensuring the probe rate is sufficiently low to avoid exciting more than one polariton in the cloud, which could result in self-blockade\,\cite{peyronel2012quantum}. Additionally, the prepared Rydberg state $\ket{75\textrm{P}_{3/2}, m_J = 3/2}$ strongly interacts with polaritons, where $C_3 = 11.15\,$GHz\,$\mu$m$^3$ for $\theta = 90^\circ$, corresponding to a large blockade radius of $\left(2C_3\sqrt{\textrm{OD}}/\gamma_\textrm{EIT}\right)^{1/3} \approx 9\,\mu$m for typical $\Omega_c/(2\pi) = 20\,$MHz used in this work. Despite the anisotropy of $D$ states, $D-P$ interactions generally show cleaner dipolar splitting and larger interaction strengths in comparison to $S-P$ interactions, particularly for the distances reported in this work. An example showcasing the purity of exchange interactions for $D-P$ states in comparison to  $S-P$ states (calculated with ``PairInteraction"~\cite{Weber2017}) is shown in Fig. \ref{DPvsSP}.

\subsection{Experimental parameters for exploring the full range of atom-polariton interaction} 

As described in the main text, measuring the full range of atom-polariton dipolar interaction as shown in Fig.~2 was obtained by varying the separation between the preparation and detection ensembles for a fixed control Rabi frequency of $\Omega_c/(2\pi) = 5.5\,\textrm{MHz}$. As the distance between the two ensembles was varied, the $480~$nm AOD frequencies were adjusted to ensure proper illumination of each ensemble during site-selective preparation and detection. The probe rate was optimized to achieve the highest photon rate while minimizing self-blockade effects, which resulted in $\Omega_p/(2\pi) \approx 130\,\textrm{kHz}$.

To improve the SNR, each data point was averaged $75$ times, obtained by performing $80$ repetitions (per MOT loading cycle) consisting of $3\,\mu\textrm{s}$ state preparation in one ensemble, $10\,\mu\textrm{s}$ switching of the control AOD frequency, followed by $10\,\mu\textrm{s}$ detection in the neighboring ensemble. Despite the long lifetime of the prepared Rydberg state ($225\,\mu$s), we find that waiting $100\,\mu\textrm{s}$ between each repetition is sufficient since the trap light expels Rydberg states.

In Fig.~2 of the main text, we plot the contrast, which is defined as the reduction in normalized transmission through the detection ensemble, as a function of the inter-ensemble dipolar interaction strength. This contrast is upper-bounded by the height of the EIT resonance,
which is $\sim 0.4$ of the normalized probe field transmission.

\subsection{Experimental parameters for investigating the building blocks of a nonlinear photonic network}

The experimental sequence of Fig.~4b, exploring the hopping of a Rydberg excitation, consists of the following steps: (1) a Rydberg excitation is prepared in the middle ensemble in $3\,\mu$s, (2) the $480\,$nm AOD frequency is switched to address the outer neighboring ensembles within $10\,\mu$s, (3) EIT detection light illuminates neighboring ensembles for $10\,\mu$s, thereby exciting Rydberg polaritons, (4) the $480\,$nm AOD frequency is switched back to address the middle ensemble within $10\,\mu$s, and finally, (5) the Rydberg population in the middle ensemble is measured via EIT detection in $10\,\mu$s, and plotted in the bar graph. Thus, the plot shows the Rydberg $\ket{n'P}$ population in the middle ensemble at the end of the sequence, depending on whether the control light is turned on in step 3 (purple) or not (green), which serves as a differential measurement of whether the generation of polaritons in neighboring ensembles affects the stationary Rydberg population. As an additional check, we perform the same experiment without loading neighboring ensembles, in which case we do not observe suppression in signal when control light is turned on in step 3, and we conclude that our observation cannot be explained by e.g., crosstalk due to finite size effects of our site-selective control beams. This confirms that the observed loss of Rydberg population is due to interactions with neighboring Rydberg polaritons. Note that for the data presented in Fig.~4c, the signal represents detection contrast normalized to the maximum contrast observed for this dataset (at $13.3\,\mu$m). For Fig.~4a, the probe rate was optimized for each control Rabi frequency separately.

\begin{figure*}[htbp]
\centering
\includegraphics[width=\textwidth,trim= 80 200 70 100,clip]{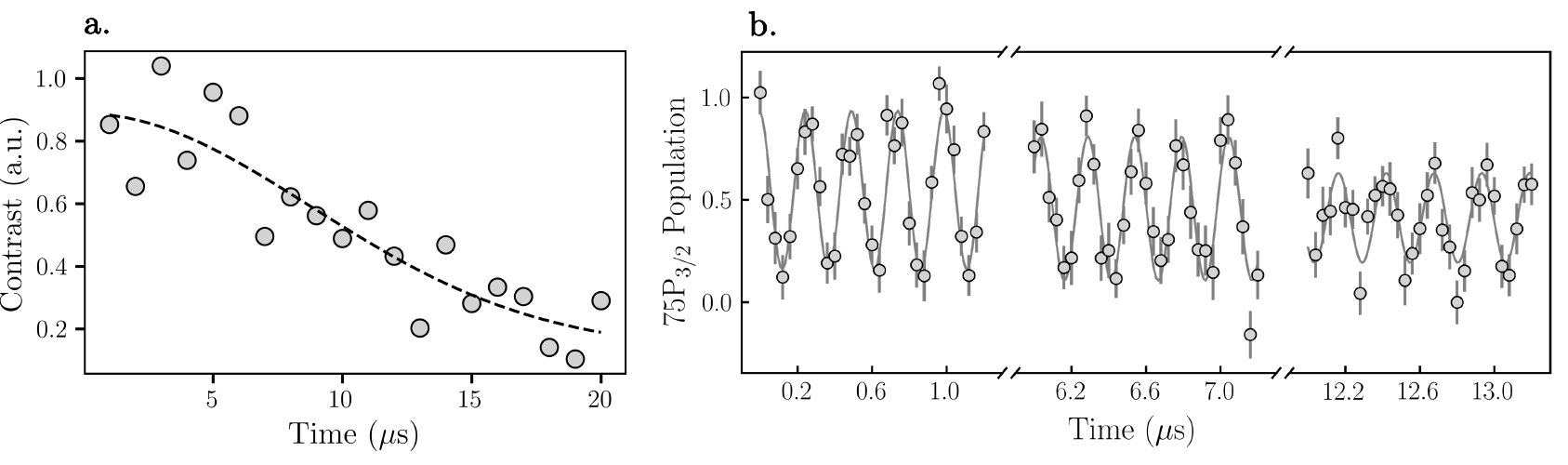}
\caption{\textbf{Coherence time and long Rabi oscillations: a.} Contrast observed in a Ramsey sequence with microwave driving of the $\ket{75\textrm{P}_{3/2}, m_J = 3/2}\leftrightarrow \ket{74\textrm{D}_{5/2}, m_J = 5/2}$ transition. From the fitted curve, we extract a coherence time of $T_2^* = 12(3)\,\mu$s. \textbf{b.} Long-range detection of Rabi oscillation shows that the prepared Rydberg atom can be driven coherently for longer than $12\,\mu$s by the microwave field.}
\label{CoherenceAndRabi}
\end{figure*}

\subsection{Microwave Rabi oscillation: experimental details}

The microwave (MW) field used for preparation and MW Rabi oscillations is a global field 
at $\nu = 5.214\,$GHz during preparation, which is $+100\,$MHz detuned from the $\ket{74\textrm{D}_{5/2}, m_J = 5/2}\,\rightarrow\,\ket{75\textrm{P}_{3/2}, m_J = 3/2}$ transition. During Rabi driving, the microwave field is on-resonance at $\nu = 5.114\,$GHz. 

We perform long-range detection of microwave Rabi oscillations in the following manner: we prepare a single Rydberg atom in the preparation ensemble in $3\,\mu$s, switch the $480 \,\textrm{nm}$ AOD frequency while applying the resonant MW field for a variable duration, and perform readout of the detection ensemble.
To achieve fast switching of the high-frequency MW field, we mix a $100\,$MHz source with a high-frequency source set to $5.214\,$GHz. The difference frequency is on-resonance with the microwave transition, which can be quickly turned off using low-frequency electronics on the $100\,$MHz source. For the data in Fig.~3c, multiple repetitions per sequence were used, as described previously.


\subsection{Rydberg coherence time}

To ensure that the prepared Rydberg  within the ensemble has a sufficiently long coherence time to interact with the neighboring Rydberg polariton, we measured the coherence time while driving the $\ket{75\textrm{P}_{3/2}, m_J = 3/2} \leftrightarrow \ket{74\textrm{D}_{5/2}, m_J = 5/2}$ transition. This was done using a Ramsey sequence, with two resonant $\pi/2$ pulses and a variable delay time in-between. The data, shown in Fig. \ref{CoherenceAndRabi}a, reveals a coherence time of $12(3)\,\mu$s, which is sufficient to allow coherent oscillations between the stationary Rydberg atom and neighboring Rydberg polaritons within a duration of the polariton lifetime (much shorter than $1\,\mu$s).

\subsection{Multiplexing ensembles for improved detection}

One unique feature of this detector compared to other free-space and cavity-based imaging methods is that multiple detectors can be employed to read out a single qubit. In the blockade regime, it is expected that each detector would observe the same signal, thus enhancing the imaging SNR by a factor $\sqrt{N}$, where $N$ is the number of detectors.

\begin{figure}[htbp]
\centering
\includegraphics[width=0.35\textwidth,trim= 270 185 390 100,clip]{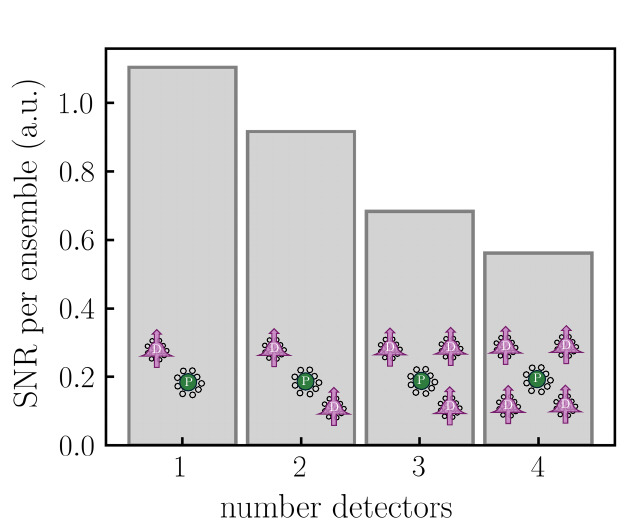}
\caption{\textbf{Multiple Rydberg detectors experiment:} Multiple ensemble detectors (purple) are prepared at a distance of $19\,\mu$m from the qubit ensemble (green). The bar graph displays the detection SNR per ensemble as a function of the number of detectors, which decreases with the addition of more detectors. However, after summing over all detectors, the total detection SNR remains comparably similar (not shown in the figure). This observation supports the occurrence of hopping at these  distances.}
\label{MultipleRydbergDetectors}
\end{figure}

Even though we were unable to reach the blockade regime in our system for optimal readout, we demonstrate the multi-ensemble detection technique as a proof-of-principle.
We prepare multiple detector ensembles equidistant from the qubit ensemble, as shown in Fig.~\ref{MultipleRydbergDetectors}, and we observe that the SNR per ensemble drops with the addition of more detectors. In other words, the total detection SNR remains comparably similar. This observation further supports the occurrence of Rydberg excitation hopping from the preparation ensemble to the detection ensembles at distances of $r = 19\,\mu$m, which lie in the probabilistic hopping regime for the control Rabi frequency $\Omega_c/(2\pi) = 20\,$MHz used in this measurement.  

\subsection{Parameters affecting detection fidelity} 

With future high-fidelity detectors in mind, we analyze the effect of each parameter on the performance of the detector and propose changes that should enable readout fidelity of $>99\%$. The most important parameters limiting detector performance, as described below, include probe photon collection efficiency, atomic optical density (OD), control Rabi frequency, Rydberg $n$ level, and the distance between the detector and the prepared Rydberg atom. 

\textit{Probe photon collection efficiency:} One simple upgrade that would enhance the detector's performance is the use of a microscope objective with better Numerical Aperture (NA) and transmission at our probe wavelength of $780\,\text{nm}$. Our current objective has $\textrm{NA} = 0.4$ and transmits $62\,\%$ of our probe photons, whereas, objectives with $\approx 90\,\%$ transmission and $\textrm{NA} = 0.65$ are readily available, which would increase our detected photon rate by a factor of $\approx 4$, thereby enhancing the SNR by a factor of $\approx 2$.

\textit{Optical depth:} We empirically determined that the optimal OD for this detector lies within the range of $0.9-1.7$ due to the competition between two factors: the probe rate, and maximum signal contrast due to probe photon absorption of the ensemble.
The number of photons detected scales with the incident probe rate, which due to self-blockade effect~\cite{peyronel2012quantum} determines that the maximal probe rate will scale inversely with OD ($\textrm{probe rate} = \Omega_{c}/\textrm{OD}$). 
However, a higher OD increases the maximum signal contrast $\textrm{C} = (1 - e^{-\textrm{OD}})$, which we define as the maximum normalized probe photon absorption. Considering both factors, the overall detection SNR scales as $\sqrt{\textrm{probe rate}} \times \textrm{C}$, which is also numerically found to be maximal in the range of $0.9-1.7$.

\textit{Control Rabi frequency:} Since the probe rate scales as $\Omega_{c}/\textrm{OD}$, higher control Rabi frequencies are preferable. However, increasing the control Rabi frequency also reduces the blockade radius as $r_b \propto \Omega_c^{-2/3}$, effectively reducing the maximum signal contrast $\textrm{C}$ since less atoms are blockaded. We empirically find that a control Rabi frequency on the order of $20~$MHz is optimal for balancing these effects.

\textit{Rydberg $n$ level:} The aforementioned two factors influence our choice of Rydberg level as well. At lower $n$, higher probe rates are permitted due to a smaller blockade radius and lower self-blockade interactions~\cite{peyronel2012quantum}. On the other hand, better detector performance is achieved in the blockade regime, which necessitates a larger blockade radius, achieved through a higher $n$. Therefore, the optimal $n$ level is the lowest one for which the interaction strength at the inter-ensemble distance is on the order of the control Rabi frequency (i.e., close to the transition point between the blockade and coherent exchange regimes, from the blockade side). Therefore, this parameter should be determined once the distance between the detector and qubit ensembles is selected.

Considering these constraints, we hypothesize that the optimal detector configuration involves positioning a large detector ensemble adjacent to a tightly-focused tweezer which confines a single-atom qubit. Assuming a $5\,\mu\text{m}$ beam waist for the detector ensemble (similar to this work) and a $\sim 1\,\mu\text{m}$ waist for the qubit, separations of $\approx 7-8\,\mu\text{m}$ can be achieved. With $\Omega_c/(2\pi) =20\,\text{MHz}$, the blockade regime can still be achieved for relatively low $n$ (e.g. 54D$_{5/2}$) at this distance. By implementing only these changes -- lower $n$ level, closer detector-qubit distance, and improved collection efficiency -- and considering a much better preparation fidelity for the single atom in a Rydberg state (approaching 100\%), we expect to achieve $\approx 97\,\%$ single-shot detection fidelity. We estimate this by scaling both the signal and noise by the appropriate factor proportional to the expected probe rate enhancement from these changes. Thus, a repeated detection like the one demonstrated in Fig.~\ref{figure3}b could, potentially, reach $>99\%$ fidelity.

To push the single-shot fidelity further, we could move the single-atom tweezer inside the detector prior to the state preparation and detection process. By preparing the qubit atom in a different hyperfine state (e.g. F = 1), we can prepare only the qubit atom in a Rydberg state (and subsequently detect it) without disturbing nearby detector atoms. This approach allows a large detector ensemble to be operated within the blockade regime. Therefore, if the detector trap radius is doubled (and assuming the number of atoms increases fourfold), the signal would increase fourfold, and the SNR would double. With these improvements, we estimate that a fidelity of $\sim 99\%$ can be achieved. In a dual-species setup, with the right choice of Rydberg levels for each species, one can predominately have van der Waals interactions instead of exchange interactions, which can further minimize hopping and improve detection fidelity~\cite{zhang2025dual}.

\subsection{Phenomenological theoretical model}

In this section, we describe the phenomenological theoretical model used to derive the fitted curve shown in Fig.~2 of the main text. Our model is based on the competition between polariton blockade and exchange mechanisms, both caused by the same dipole-dipole interaction term. The contrast, i.e. normalized amount of absorbed photons, is described by the equation \( C = p_{\textrm{block}} \, \cdot \, C_{\textrm{block}} + (1 - p_{\textrm{block}})\, \cdot\, C_{\textrm{hop}} \), where \( C_{\textrm{block}} \) is the contrast expected from blockade of polariton formation given the atom-polariton interaction strength and the EIT linewidth, \( C_{\textrm{hop}} \) is the contrast signal expected from hopping, and \( p_{\textrm{block}} \) is the probability that a polariton will be blocked. The probability \( p_{\textrm{block}} \) is further defined as \( p_{\textrm{block}} = (\frac{V_\textrm{PD}(r)} {\gamma_\textrm{EIT}})^2 \), 
where for distances \( =V_\textrm{PD}(r) \gt \gamma_\textrm{EIT} \) (in the blockade regime), \( p_{\textrm{block}} \) is set to 1.


The blockade contrast, $C_{\textrm{block}}$, is defined as the constant contrast due to blockade of polariton formation, which is a free parameter in the theory curve, found to be approximately $0.23$. This value coincides with the saturated contrast observed in Fig.~2 of the main text for large control Rabi values. 

The hopping contrast, $C_{\textrm{hop}}$ is a bit more involved, and is defined in the following way: We assume that the polariton is present inside the cloud over a duration given by the polariton lifetime \( \tau_d = \textit{OD}\frac{ \Gamma }{\Omega_c^2} \), during which the polariton and Rydberg atom interact with a strength of $V_\textrm{PD}(r)/2 = C_3 / r^3$. When the polariton exits the ensemble, the Rydberg atom and the polariton are described by the entangled state $\Psi_{\textrm{polariton, atom}}\, =\, \alpha\ket{DP} + \beta\ket{PD}$. The polariton can then exit the ensemble as a transmitted photon, which is then detected by the camera with a probability of \( \alpha^2 \), projecting the detection ensemble into a state with no Rydberg excitations for the next incident photon. Alternatively, the polariton can be projected into the \( P \) state with a probability of \( \beta^2 \), which causes the next incident photon, and all subsequent photons during the detection period, to be blocked. This is a crucial element in our model, which highlights the state memory inherent to the system. Thus, our theory assumes that until the \( P \) Rydberg excitation has hopped into the detector ensemble, the contrast in that ensemble is zero (i.e. photons are transmitted). Once the \( P \) excitation has hopped, the contrast in that ensemble will be a constant, given by the blockade level contrast, as the local \( P \) state will block subsequent photons from entering. The probability that the \( P \) state is projected into the polariton cloud is given by \( \beta^2 = \frac{1 - \cos\left(2\pi \cdot V_\textrm{PD} \tau_d\right)}{2} \). 
We also assume that the hopping between the polariton and the Rydberg atom is imperfect and that the contrast decays with rate \( \Gamma_d \). This is expressed as \( \beta^2 = \frac{1 - e^{-\Gamma_d \tau_d} \cos\left(2\pi \cdot V_\textrm{PD} \tau_d\right)}{2} \). 
By combining the above probability with the probe rate, we calculate the duration it takes for the \( P \) state to hop within the EIT detection duration of \( 10\,\mu\text{s} \) used in this measurement. The hopping contrast, \( C_{\textrm{hop}} \), is then equal to the fraction of time the \( P \) state was present in the detector ensemble. We note here that we also take into account that the Rydberg fraction of the polariton is not unity, by multiplying the fraction with the probability above. 

Using this model, we fit the data with free parameters that ultimately match the experimentally estimated values. 
The control Rabi frequency is fitted to be $ \Omega_c/(2\pi) = (4.8\, \pm \, 0.1) \, \text{MHz} $ (experimental value $ \Omega_c/(2\pi) = 5.5 \, \text{MHz} $), and the optical depth (OD) is fitted to be $ \text{OD} = 1.15\, \pm \, 0.10 $ (experimentally observed peak OD is approximately $2$, with an average OD of approximately $1.35$).

\subsection{Ab-initio theory of transitions in Rydberg atom-polariton dipolar interaction}

While the description of the stationary Rydberg atom is simple, and consists only of two states ($\ket{nD}$ and $\ket{n'P}$), the description of the collective excitation of atoms in the detector ensemble seems, at face value, rather complicated. However, since the collective excitation is coherent, there is a compact form to write the state and the Hamiltonian of the system, within the rotating wave approximation and in the rotating frame, when one assumes a single excitation in the ensemble~\cite{thompson_symmetry_protected,bienias2014scattering}:

\begin{widetext}
\begin{equation}
\cal{H}_{\text{single}} = \hbar\int 
\begin{pmatrix}
\psi_\textrm{E} \\
\psi_\textrm{I} \\
\psi_\textrm{D} \\
\psi_\textrm{P} \\
\end{pmatrix}^{\dagger}    
\begin{pmatrix}
-i c \frac{\partial}{\partial z} & \Omega_p & 0 & 0 \\
\Omega_p & -i \Gamma & \Omega_c & 0 \\
0 & \Omega_c & -i \Gamma_\textrm{D} & 0 \\
0 & 0 & 0 & -i \Gamma_\textrm{P}
\end{pmatrix}
\begin{pmatrix}
\psi_\textrm{E} \\
\psi_\textrm{I} \\
\psi_\textrm{D} \\
\psi_\textrm{P} \\
\end{pmatrix}\,dz
    \label{Eq:single}
\end{equation}
\end{widetext}
where the notations $\psi_{E}, \psi_{I}, \psi_{D}$ and $\psi_{P}$ denote the bosonic operators of a single photon (ensemble in the ground state), a single excitation in the intermediate $\ket{5P_{3/2}}$ state, a single $\ket{nD}$ excitation in the ensemble, and a single $\ket{n'P}$ excitation in the ensemble, respectively. $-i c \frac{\partial}{\partial z}$ represents the photon propagation kinetic energy; $\Gamma, \Gamma_\textrm{D}$ and $\Gamma_\textrm{P}$ are the decay of the $\ket{5P_{3/2}}$, $\ket{nD}$ and $\ket{n'P}$ states, respectively; and $\Omega_p, \Omega_c$ are the Rabi frequencies of the probe and control fields. 

In our system, electromagnetically induced transparency occurs between the $\ket{5P_{3/2}}$ state and the $\ket{nD}$ state, and thus the $\ket{n'P}$ state is unreachable without external influence on the polariton. The lifetime of the $\ket{nD}$ and $\ket{n'P}$ states is far larger than the timescales in our experiment, allowing us to neglect them, and we also neglect the contribution of the photon propagation since it only accounts for a slight deviation from the EIT condition. The single-excitation assumption is justified since the experimental conditions were below the self-blockade conditions, with the van der Waals interaction preventing any further excitations within the ensemble. Therefore, we can describe the system of the interacting atom and polariton via an effective two-body Hamiltonian of the following form:

\begin{widetext}
    \begin{equation} 
    \cal{H}_{\text{full}} = \hbar
    \begin{pmatrix}
    \psi_\textrm{E1D2} \\
    \psi_\textrm{E1P2} \\
    \psi_\textrm{I1D2} \\
    \psi_\textrm{I1P2} \\
    \psi_\textrm{D1D2} \\
    \psi_\textrm{D1P2} \\
    \psi_\textrm{P1D2} \\
    \psi_\textrm{P1P2}
    \end{pmatrix}^{\dagger}
    \begin{pmatrix}
    0 & 0 & \Omega_p & 0 & 0 & 0 & 0 & 0 \\
     0 & 0 & 0 & \Omega_p & 0 & 0 & 0 & 0 \\
     \Omega_p & 0 & -i \Gamma & 0 & \Omega_c  & 0 & 0 & 0 \\
     0 & \Omega_p & 0 & -i \Gamma & 0 & \Omega_c  & 0 & 0 \\
     0 & 0 & \Omega_c  & 0 & 0 & 0 & 0 & 0 \\
     0 & 0 & 0 & \Omega_c  & 0 & 0 & V_\textrm{PD} & 0 \\
     0 & 0 & 0 & 0 & 0 & V_\textrm{PD} & 0 & 0 \\
     0 & 0 & 0 & 0 & 0 & 0 & 0 & 0 \\
    \end{pmatrix}
    \begin{pmatrix}
    \psi_\textrm{E1D2} \\
    \psi_\textrm{E1P2} \\
    \psi_\textrm{I1D2} \\
    \psi_\textrm{I1P2} \\
    \psi_\textrm{D1D2} \\
    \psi_\textrm{D1P2} \\
    \psi_\textrm{P1D2} \\
    \psi_\textrm{P1P2}
    \end{pmatrix}
    \label{Eq:full}
    \end{equation}    
\end{widetext}

where the numbers 1 and 2 denote the polariton operator in the detection ensemble and the atom operator in the preparation ensemble, respectively. $V_\textrm{PD}$ is the dipole-dipole exchange interaction term and it is possible to neglect all other interaction terms in our system due to the distance between the detector and preparation ensembles. The system in Eq. \ref{Eq:full} can be rearranged into a block-diagonal form, with an interacting block and a non-interacting block. Then, we can consider only the compact interaction Hamiltonian:

\begin{widetext}
\begin{equation}
    \cal{H}_{\text{compact}} = \hbar
    \begin{pmatrix}
    \psi_\textrm{E1P2} \\
    \psi_\textrm{I1P2} \\
    \psi_\textrm{D1P2} \\
    \psi_\textrm{P1D2} \\
    \end{pmatrix}^{\dagger}    
    \begin{pmatrix}
    0 & \Omega_p & 0 & 0 \\
    \Omega_p & -i \Gamma & \Omega_c & 0 \\
    0 & \Omega_c & 0 & V_\textrm{PD} \\
    0 & 0 & V_\textrm{PD} & 0
    \end{pmatrix}
    \begin{pmatrix}
    \psi_\textrm{E1P2} \\
    \psi_\textrm{I1P2} \\
    \psi_\textrm{D1P2} \\
    \psi_\textrm{P1D2} \\
\end{pmatrix}
    \label{Eq:compact}
\end{equation}
\end{widetext}

We can now diagonalize the Hamiltonian of Eq. \ref{Eq:compact}, and consider the two eigenstates affected by the interaction term $V_\textrm{PD}$. Fig. \ref{figure2}b of the main text shows the change in these two eigen values, as a function of $V_\textrm{PD}$, given the experimental parameters in our system. A critical point is clearly visible, where the eigenvalues shift from being completely imaginary (large $V_\textrm{PD}$) to having a complex value with a considerable real part (small $V_\textrm{PD}$). The result is in-line with our understanding that polaritons decay inside the ensemble for large interaction strengths, which inhibit polariton formation; and produce two interaction-induced dressed states for small interaction strengths, which allow the polaritons to form. 

\begin{figure*}[htbp]
\centering
\includegraphics[width=0.35\textwidth,trim= 0 0 0 0,clip]{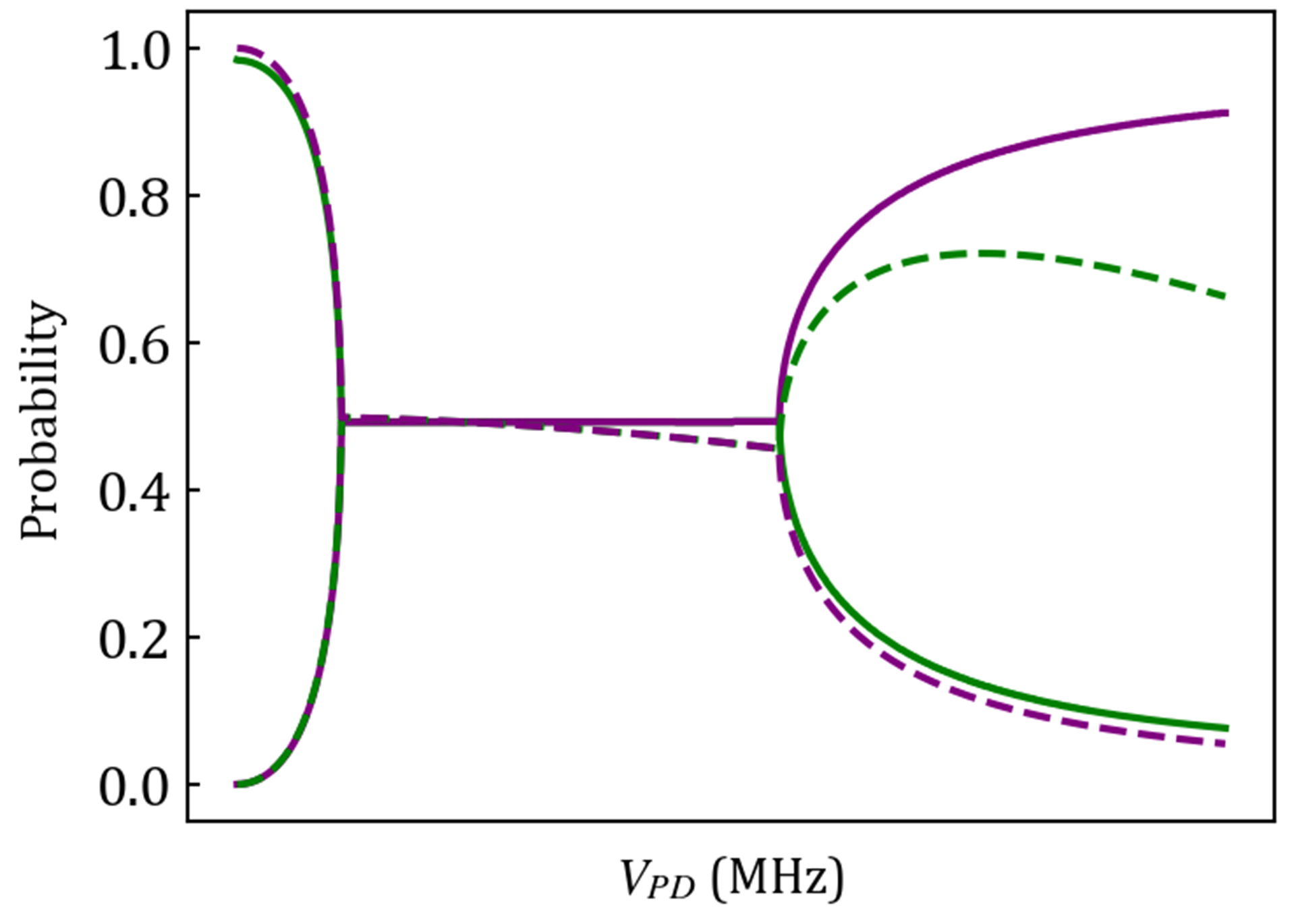}
\caption{\textbf{Artificially reproducing the transition between coherent exchange and incoherent hopping:} The probability to be in the states $\ket{\psi_{E1P2}}$ (solid lines; a photon in the detection ensemble) and $\ket{\psi_{P1D2}}$ (dashed lines; an $\ket{n'P}$ state in the detection ensemble), for the two polariton eigenstates (green and purple). The addition of an $\ket{nD}$ state lifetime changes the eigenstates at low interaction energies, such that one eigenstate is a pure $\ket{n'P}$ state in the detector ensemble, and the other is a nearly pure photon state. This means that any result of polariton propagation in this regime acts as a classical rate equation, where the end result is, to a very good approximation, either an outgoing photon or a stationary $\ket{n'P}$ atom in the detection ensemble. This result captures the essence of the transition between the coherent exchange and incoherent hopping regimes, yet since the observation here is merely qualitative, we do not give interaction strength values, as they do not reflect the values measured in experiment.}
\label{phasetransition_D_lifetime}
\end{figure*}

The sudden change in eigenvalues, where an energy gap opens between the two polariton states, is indicative of an exceptional point~\cite{heiss2012physics} being crossed in the transition between the blockade and coherent exchange regimes, which is usually an unequivocal evidence of a dissipative phase transition, when occuring in a many-body system~\cite{kessler2012dissipative,Fitzpatrick2017Observation,fink2018signatures,PhysRevLett.134.130404}. The exceptional point persists even in the limit of an infinite number of atoms in the ensemble, albeit for different numerical values, further supporting the notion of a phase transition. However, there is still ambiguity if a dissipative phase transition is the correct definition of this phenomenon, for although it occurs in a many-body system, it is in fact limited to exhibiting only effective two-body dynamics, due to the hard-core boson nature of the Rydberg excitations. There is also some similarity in this transition to PT-symmetry-breaking transitions~\cite{Bender2024PT}, yet it is unclear how to define parity in such an imbalanced system.

Intriguingly, our system bears great similarity to an open central spin system~\cite{kessler2012dissipative} - a model system exhibiting a dissipative phase transition - since the Rydberg atom interacting with a Rydberg polariton in an adjacent ensemble is akin to an ensemble of non-interacting spin-1/2 particles uniformly coupled to a single spin-1/2 particle. The similarity to the open central spin system persists when considering the Hamiltonian, as both have the self-energies of the single particle and the ensemble, as well as a dipolar interaction term. Finally, the order parameter for the phase transition in the open central spin system is the ratio of the driving frequency to the interaction strength, which has the same role as $V_\textrm{PD}$ in our system, when normalized by the control Rabi frequency. The major differences between the models are the hard-core boson nature of our system, which precludes the development of a large manifold; the driving and dissipation being applied to the ensemble in our case, instead of the central spin; and the negligible strength of van der Waals interaction in our system due to the atom-polariton distance.

Within our Hamiltonian description above there are no time-dependent interactions or contributions from the photon group delay inside the ensemble. The latter produces another competition between timescales, as the atom-polariton interaction must be faster than the travel time of a photon in the ensemble if it is to remain coherent, giving rise to the transition between coherent exchange and incoherent hopping at the hopping radius $r_h$. Although our description cannot strictly account for this phenomenon, we can effectively recover it by introducing an artificial lifetime for the $\ket{nD}$ state. Fig. \ref{phasetransition_D_lifetime} shows qualitatively how another transition occurs due to this finite lifetime, and the new eigenstates of the system at very low interaction strengths represent the rate equation limit: the detector ensemble probability is either fully in the outgoing photon or in a stationary $\ket{n'P}$ state. 

However, this description does not quantitatively agree with our experimental values, since this lifetime was artificially inserted, and a full time-dependent treatment, beyond the scope of this manuscript, is necessary to obtain the correct energy for this second transition. At any rate, this transition is the result of a finite system size and therefore cannot strictly be called a phase transition. It also holds a relatively trivial analogy to a pair of coupled waveguides, offset in their decay rates~\cite{el2018non}, which is a single-body phenomenon.

\clearpage

\end{document}